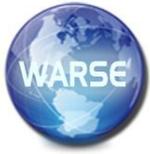

# Using Chaotic Stream Cipher to Enhance Data Hiding in Digital Images

**Sana' Haimour[1], Mohammad Rasmi AL-Mousa[2], Rashiq R. Marie[3]**
[1]Department of Computer Science, Zarqa University, Zarqa, 13132, Jordan, soso_haimour_89@yahoo.com
[2]Department of Software Engineering, Zarqa University, Zarqa, 13132, Jordan, mmousa@zu.edu.jo
[3] Information System Department, Taibah University, Madinah, Saudi Arabia, rashiqrafiq@gmail.com

**ABSTRACT**

The growing potential of modern communications needs the use of secure means to protect information from unauthorized access and use during transmission. In general, encryption a message using cryptography techniques and then hidden a message with a steganography methods provides an additional layer of protection. Furthermore, using these combination reduces the chance of finding the hidden message. This paper proposed a system which combines schemes of cryptography with steganography for hiding secret messages and to add more complexity for steganography. The proposed system secret message encoded with chaotic stream cipher and afterwards the encoded data is hidden behind an RGB or Gray cover image by modifying the kth least significant bits (k-LSB) of cover image pixels. The resultant stego-image less distorters. After which can be used by the recipient to extract that bit-plane of the image. In fact, the schemes of encryption/decryption and embedding/ extracting in the proposed system depends upon two shred secret keys between the sender and the receiver. An experiment shows that using an unauthorized secret keys between the sender and the receiver have totally different messages from the original ones which improve the confidentiality of the images.

**Key words :** Chaotic Stream Cipher, Data Hiding, encryption, decryption, Steganography.

## 1. INTRODUCTION

Cryptography refers to the study and examination of coded (not understood) methods to deliver messages, in order for the concerned recipients only to read the message and have the disguise removed. Through turning it into an unreadable format, it preserves information [1, 23]. Nevertheless, numerous people have been joining the cyberspace movement, steganography becomes more relevant. The importance of concealing information lies in improving the level of security when exchanging information, as it can be used in a different area of information security, such as an encryption system, and for preserving evidence in the network forensic [29-31]. Steganography refers to the art of having information concealed in a manner that prevents the messages that are hidden from being detected [23]. Steganography aims to avoid having suspicion about the existence of a message that is hidden. In general, the main categories of information hiding are separated into two classes; Steganography and Digital Watermarking as shown in Figure 1.

The aim of this paper is to design and implement a hiding system of information to inject a secret message into an image media file. Consequently, upon insertion of the document, no material changes to the image file will occur. At the same time the embedded message is encrypted prior embedding using chaotic-based cryptography to increase the level of security.

Steganography refers to the art of having details disguised in a manner that shall prevent the secret massages from getting detected. In addition, it is a science of communication in a manner which hides the distance that covers the communication life. It refers to the art of having information transmitted in a manner that doesn't know the very life of the post, and hides a secret message within a wider one so that others are unable to detect the existence or meaning of the hidden message [2-5].

Steganography aims at delivering messages in a safe and undetectable manner. It aims to avoid having the attention drawn to the delivery of confidential data. [6, 7]. It prevents others from believing that there is information being delivered. Steganography aims at avoiding having attention drawn towards the delivery of hidden messages.

If suspicion arise, it means that the goals of steganography wasn't met. If a steganography method causes someone to suspect the carrier medium, then the method has failed [8].

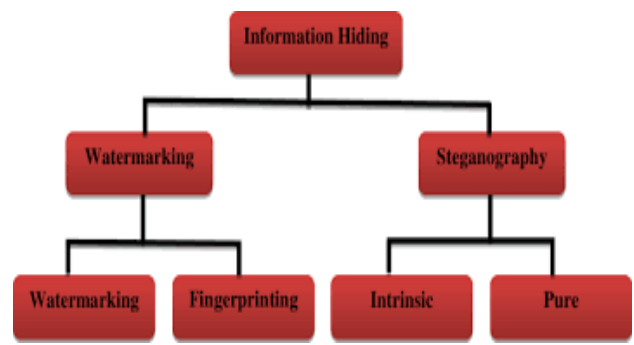

**Figure 1:** Classification of information hiding

Figure 2 shows the basic model of steganography. This model involves a carrier, a password and a massage. Carrier may be called the cover-object, which includes the message. It provides assistance in hiding the existence of a message.

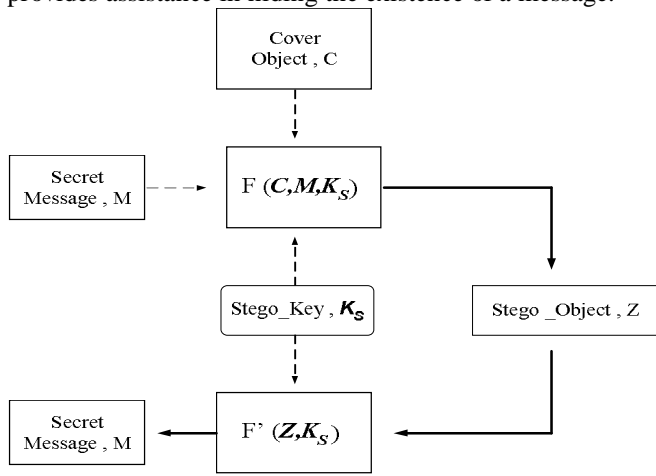

**Figure 2:** The basic model of steganography.

The relation of encoder and decoder of a stegosystem represented, by as the following [10]:

$$Z = f(C, M, K_s) \text{ and } M = f^{-1}(Z, K_s) \qquad (1)$$

Where Z is the stego-object, C is the cover-object, M is the message, and Ks is the stego-key.

Within various media types, secret message can be hidden. Nowadays, most steganography hide information within images. Doing that is easy.

Regarding the most important property of the cover source, it's represented in the amount of data which can be hidden. If the image got distorted or a piece of music sound became different from the original sound, the cover source shall be deemed as suspicious. In this case, the messages shall become suspicious and checked.

The image being pieces of data that most frequently exchanged over the internet. An advantage in using images for data hiding is they represent an innocent medium, since it is possible to access any pixel of the image at random.

Usually, digital color images get stored in 24-bit formats. Regarding the use of the color format of RGB, it is called (true color). Regarding all the color variations of the pixels of the image which format is 24-bit, they are created through using 3 main colors: green, red and blue, with 8 bits representing each primary color. Therefore, there can be 256 different amounts of red, green, and blue in a single pixel, adding up to more than sixteen million variations. That shall lead to having more than sixteen million colors. The larger the amount of colors which may be presented, the larger the size of the file shall be [14], [15], [16].

Usually, spread spectrum communication is ideal used in Radio Frequency (RF) communication as it main component is noise. However, spread spectrum in image treat the cover image as noise or through having the pseudo-noise added to the cover image [7]. There are three keys used in the SS (Spread Spectrum) method. The keys are used for the message encryption for seed the pseudo-random noise generator and to interleave the secret data in order to spread it over the image. There are steganography techniques. Such techniques include the following:
- Steganography of spread spectrum
- Statistical steganography,
- Steganography of Cover generation

Through the section below, the researcher sheds a light on the description of the LSB because it is the selected technique in the project.

## 2. CRYPTOSYSTEM BASED ON CHAOTIC SYSTEM

Regarding the Chaotic systems, they have several significant properties. Such properties include: having sensitive dependence on system parameters and initial conditions. These parameters suit strong cryptosystem requirements. Recently a growing number of researchers have been working in this area, resulting in a variety of cryptosystem designs based on chaotic systems.

A cryptosystem (cipher) is considered an algorithm that converts the plaintext (i.e. original message) into a ciphertext (i.e. scrambled message). It turns the message into its initial form. The transformation of a message from plaintext into ciphertext is regulated by a key. This key is called (encryption). The transformation from ciphertext into plaintext is regulated by a key. This key is called (decryption). Denote the plaintext and the ciphertext by P and C, respectively. The procedure of the encryption of a cipher may be described as follow $C = E(P, K_e)$:

Whereas $K_e$ refers the encryption key and E ( ) is the encryption function. Regarding the procedure of the decryption of a cipher may be described as follows: $P = D(C, K_d)$, see Figure 3.

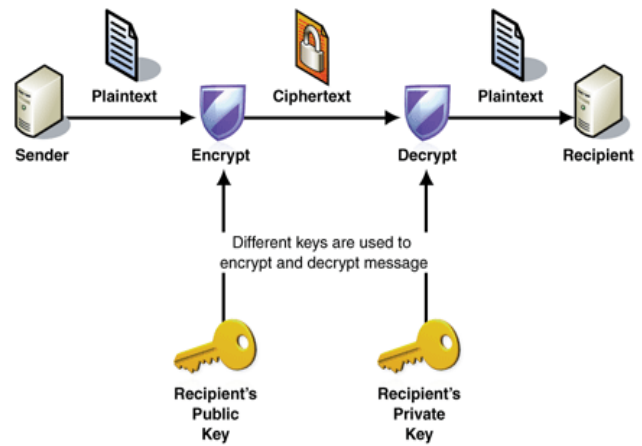

**Figure 3:** Encryption-Decryption Process [28]

### 2.1 Symmetric and Asymmetric Cipher

There are two types of ciphers that fill $K_e$ and $K_d$ relationship, the cipher is referred to as a private-key cipher or a symmetric cipher when $K_e = K_d$(Crypto-Key). Regarding the key of encryption, it must get transmitted from the sender to the recipient through an encrypted special channel for private-key ciphers.

The cipher may be named a public-key or asymmetric cipher when $K_e$ not eqaul $K_d$. The encryption key $K_e$ is released for public –key ciphers, and the key of decryption $K_d$ is kept private for which there isn't any additional secret channel needed for the transfer of the key.

Regarding the main problem with symmetric key cryptosystems, it lies in the secret key k which shall be shared between the sender and the recipient. Nevertheless, one could overcome that problem through sharing the secret key through employing a secure channel or through employing asymmetric cryptosystem.

### 2.2 Block and Stream Ciphers

Ciphers can be divided into two groups in accordance with the encryption structure.

This structure consists from stream ciphers and block ciphers. Block ciphers encrypt block by block the plain text block, and mapping each block with the same size into another block. Flow ciphers encrypt the plain text through employing the encryption key powered by a pseudo-random (called key-stream). Regarding the binary additive stream cipher, it is a synchronous stream cipher in which binary sequences are the key-stream, plaintext and ciphertext.

### 2.3 Chaotic Systems

Chaotic refer to the systems that are nonlinear and show a random behavior for a specific set of system parameter values. Now, there is a great number of chaotic systems available, both physical and mathematical, which could theoretically act as hardware and software for encryption and decryption. The nonlinear simple systems which obey iterative dynamics are considered as potential generators of dynamics that are complicated, as mentioned in [22]. It is this complexity that takes on significance in cryptographic encryption / decryption algorithms.

### 2.4 Properties of Chaotic Systems

In the light of the initial phase of the deterministic system; in general, the nonlinear system, it is well understood that predication may be made on the future states of the system. Long-term prediction is however difficult for chaotic systems. On initial conditions, the chaotic system is reactive. Characterize chaotic dynamics by the following properties. The significant features of instability are its intense vulnerability to the system's initial conditions. Slight difference in the direction of the initial conditions causes completely difference. Two trajectories, which at first are very similar, diverge in an exponential manner within a short duration for different parameter values.

### 2.5 Characteristics of the Chaotic Maps

Most properties of the chaotic maps relate to some requirements in the sense of cryptography, such as mixing and diffusion. Chaotic cryptosystems thus have more practical and useful applications.

Regarding the logistic map, it is a simple mathematical model. It shows complex bewildering behavior. It's a simple form of the chaotic process. However, it is often employed to describe the biological population growth [22].

Later [24] the logistic map was studied as the generator of pseudorandom numbers, and is provided through:

$$X_{n+1} = F(X_n) = \lambda X_n (1 - X_n), ......$$
$$where \quad .. \quad ..X_n \in (0,1) \quad and \quad ...\lambda \in (3.96, 4]$$

Where; $X_n$ and $\lambda$ are the system variable and parameter, respectively. *n* is the number of iterations. Based on initial value $x_0$ and a parameter $\lambda$; the series $\{X_n\}$ is measred. In the present project, for meeting the goals of the chaotic cryptosystem, the researcher refer to $x_0$ and $\lambda$ as the initial state of the logistic map [24].

Figure 4 as suggested by Ref. [24] displays the bifurcation diagram of that map. That's a plot of the logistic map as a function of $\lambda$. For $0 \leq \lambda \leq 1$; the trivial solution is considered as the only point that is fixed. For $1 \leq \lambda \leq 3$; the researcher have a non-trivial fixed point. For $3 \leq \lambda \leq 3.57$, the map shows the phenomenon of the periodic doubling. For $3.57 < \lambda \leq 4$, the map became chaotic.

At last, the researcher found that the chaos values get generated in the complete range between 0 and 1. In addition, it's presumed that the above map has good qualities as a pseudorandom number generator when $\lambda \approx 4$

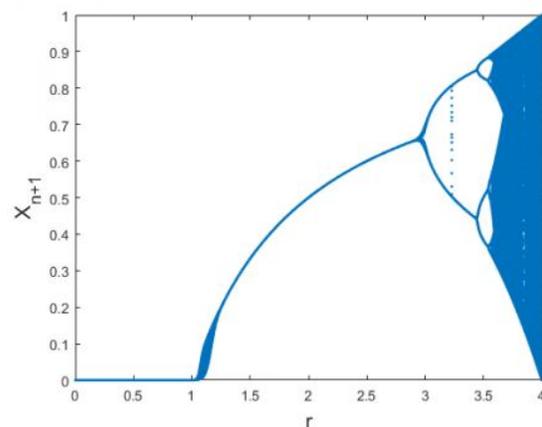

**Figure 4:** Bifurcation diagram for the Logistic Map [24]

The logistic map for binary key stream generation for cryptographic applications as mentioned in [24], is of the form $X_{n+1} = 3.999 X_n (1 - X_n), X_n \in (0,1), X_0 \in (0,1)$, i.e. we fixed the value of $\lambda$.

### 2.6 Chaotic Encryption Technique

In 1998, Baptista proposed a method of Chaotic Encryption. This method is better than the methods of traditional encryption which are employed today. Regarding the chaotic encryption, it's considered as trends in the field of cryptography. It employs chaotic system properties. Such properties include: sensitivity to the initial condition. They including having much information [25].

The chaotic systems are considered sensitive to initial conditions. They are sensitive to system parameters. When it comes to a specific set of parameters within a chaotic regime, 2 initial close conditions shall lead the system into trajectories that are divergent. Therefore, the scheme of encryption / decryption may be obtained in case the parameters are selected as "Keys" and "Trajectories" get employed for encryption/decryption [26-27].

The chaos scheme is symmetrical. The same parameters are employed for encryption and decryption. The initial conditions and parameters form a large key space that shall improve the protection of the code.

#### 2.6.1 Chaotic stream cipher

A common method for building a stream cipher is represented in employing pseudorandom sequence generator and mask the plaintext. That can be done through employing the output of the keystream generator in the aim of having the cipher text produced. Several stream ciphers were proposed in the relevant literature [21, 24]

In this project, the researcher employed a binary sequence generators. He employed them based on the chaotic Logistic map. This function is examined for generating pseudorandom binary keystream for stream cipher application. Regarding the encryption step, it consists from a simple bitwise XOR operation of the plaintext binary sequence with the keystream binary sequence to have the sequence of ciphertext binary produced [24].

To decrypt the message, at the receiver side, The Logistic Map is iterated with the same initial condition as much times as indicated by the Ciphertext to produce the same pseudorandom bit stream the inverse operation (another "xor") is done. . The encryption / decryption key (crypto-key) is represented by the initial condition $x_0$.

#### 2.6.2 Chaotic pseudorandom bits generator

The generation of a pseudorandom binary sequence from the orbit of the logistic map requires mapping the system's state. Look at Figure 5. A simple method to turn a real number $\{X_n\}$ into a discrete bit symbol $\{Z_i\}$ is represented in employing a threshold function [24]:

$$Z_i = F(X_i) = \begin{cases} 0, if\ X_i \prec c \\ 1, x_i \geq c \end{cases}$$

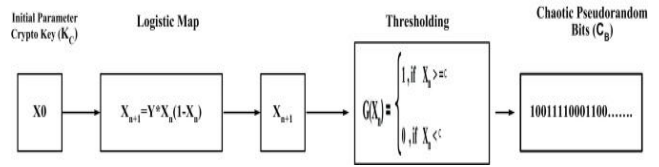

**Figure 5:** block diagram of Pseudo Random Bit Generator (PRBG)

Here, c is an appropriately chosen threshold value for $x_i$: For balanced binary sequence $\{z_i\}_{i=0}^{\infty}$ should be selected such that the likelihood of $X_i < c$ is equal to that of $X_i \geq c$

As mentioned by [24], it is proved that, for $\lambda \approx 4$, the logistic map is ergodic, which implies that $\forall i$ and almost $\forall X_0$:

1. $z_i$ and $z_{i+1}$ behave as if they are independent statistically,
2. $z_i$ is likely to be 0 or 1 equally.

The two symbols 0 and 1 are equally likely to occur for almost all $x_0$; it is also true that, for any positive integer *m*, all *m*-bit strings occur with equal probability. This is the primary reason behind selecting the parameter $\lambda$ =3.9996. For the logistic map, 0.5 is approximately the middle of the maximum and minimum of the $x_i$ values, thus, c =0.5 shall be a deemed as a perfect choice [24].

### 3. DEVELOPED CRYPTSTEG SOFTWARE

In this project, two layers of security will be used to secure the hidden information and to add more complexity for steganalysis. The information will be hidden in a cover image through employing the combines schemes of cryptography and Steganography in one system called CryptSteg .Then the stego-image will be sent via insecure channel to the receiver who will retrieve the hidden information using inverse steganography [23].

Regarding the steganography, it is not intended for it to replace cryptography. In fact, it is intended for the steganography to supplement the cryptography. If a message got encrypted and hidden through employing the steganographic method, it shall provide an additional layer of protection. It shall reduce the probabilities of having the hidden message detected. In fact, the system of CryptSteg consists from 2 primary processes [23].

- **Embedding the secret message in a cover Image;** by firstly scramble the contents of a message, chaotic encryption then canceling the existence of the message, and
- **Extracting the secret message from the stego-image;** by firstly retrieving the data then unscramble it.

## 3.1 Embedding the Secret Message

By this process the sender who wishing to send a *secret message* to the receiver will conceal the existence of this message through embedding it in the *cover image*. However, the hiding process consists of two algorithms: *Encryption algorithm* and *embedding algorithm.*

### 3.1.1 Encryption Algorithm

**Input:** Plaintext Message, Crypto-key
**Output:** Ciphertext Message
The steps of the encryption algorithms are as follows:
**Step1:** Read the plaintext ( secret message ) and convert it to a stream of binary bits ( 0's and 1's) , say,$\{X_n:$ n=1,2,3,......,N$\}$, where N number of bits.
**Step2:** Choose a secret key (Crypto Key), say $K_c$.
**Step3:** Generate a sequence of a pseudo chaotic random numbers of size N using the Logistic Function described in Chapter 4, say, $\{Y_i: i=1,2,3,....N\}$.
**Step4:** Generate a stream of binary bits, say $\{Z_n: n=1, 2, 3…N\}$, by applying the threesholding technique on the resulted sequence.
**Step5:** Apply the XOR operation between both the generated binary bits in step1 and Step 4, to get a stream of binary bits $\{Cn: n=1, 2, 3, …, N\}$
$C_n = X_n$ xor $Z_n$    *(N: number of bits).*
**Step6:** Covert the sequence of binary bits resulted in step5, and convert it characters to produce the cipher text, which will then forward to the Embedding Algorithm. Figure 6 shows the block diagram of encryption algorithm.

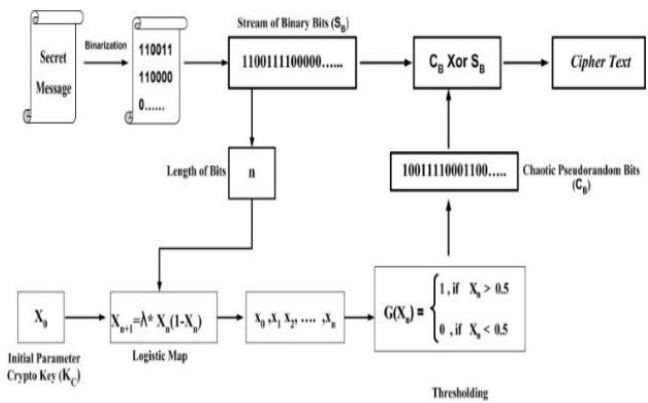

**Figure 6:** Block Diagram of Stream Chaotic Encryption

### 3.1.2 Embedding Algorithm

The embedding algorithm takes the encrypted message and scatters it randomly throughout a selected *cover- image* and produce the *stego-image*.
**Input:** encrypted secret message, Cover Image, Stego-key
**Output:** Stego-image
The following algorithm presents the steps of the process:
**Step1:** Read the encrypted form of the secret message and convert it to a stream of binary bits.
**Step2:** Generate a secret key (*Stego-key*), $K_s$, for embedding process.
**Step3:** Use the stego-key to be the seed to generate a sequence, say $\{R_n\}$, of integer random numbers of size N.
**Step4:** Choose a Cover image, say $I_{wxh}$, where *w* and *h* are image's width and height, respectively.
**Step5:** Embed the secret message in the pixels of the cover image according to the values obtained in Step3 to locate the positions of pixels in the cover image for the purpose of embedding.
**Step6:** Use the Least Significant Bit (LSB) technique to hide the binary bits of the Cipher text in the cover image.
Note that the generated *Stego-key* in step2 above is depending on the secret text, so each time we run the system and changing the secret text then the *Stego-key* will changed.
Also in Step5 the sender have options of use *k*-LSB, where k=1, 2, 3, 4 (1LSB, 2LSB, 3LSB or 4LSB). That's mean when he/she choose 1-LSB then he/she can embed one bit of the secret message in a pixel of the cover image, and similarly if he/she choose 2LSB then he/she can embed 2 bits in a pixel of the *cover image* and so on.
It is necessarily for the receiver to know what is number of LSBs was used by the sender in embedding the secret message for each pixel in the cover image. In fact the choosing of any of LSB options depend upon a previous agreement between the sender and the receiver.
In this process the receiver will try to get the recovered text *(secret text)* form the received *stego-image* and this will accomplish through two the algorithms, *Extracting Algorithm* and *Decryption Algorithm*, as follows:

### 3.1.3 Extracting Algorithm

In fact, this algorithm is a reverse order of the embedding algorithm. The process of extracting the secret information from the stego-image $S_{nxm}$ is as follows:
**Input**: stego-image, Stego-key
**Output:** encrypted secret message
**Step1:** Read the *stego-image, $S_{W*H}$*
**Step2:** Enter the *Stego-key*, $K_s$, which is the same key that was used in embedding algorithm to generate a sequence integer random numbers.
**Step3:** Extract the secret message from the *stego-image* following the random sequence of random numbers generated in step 2 to locate the positions of the *stego-image* pixels that was used in embedding process.
**Step4:** Depend upon a previous agreement one of the of LSB's types, i.e., 1LSB, 2LSB, 3LSB or 4LSB.
**Step5:** Concatenate the retrieved binary bits from the Stego-image.
**Step6:** Convert the resultant stream of bits of Step *5* to the corresponding characters which will form the cipher text of the secret message, which then forward to the decryption process.

### 3.1.4 Decryption Algorithm

The decryption process operates as follows:
**Input**: Encrypted secret message, Crypto-key
**Output**: Plain text of the secret message
**Step1**: Read the Ciphertext of the secret message obtained from the extracting algorithm and convert it to a stream of binary bits ( 0's and 1's) , say,$\{X_n: n=1,2,3,\ldots,N\}$.
**Step2**: Enter the secret key (Crypto Key), $K_c$, .The same key that was used in the encryption algorithm.
**Step3**: Generate a sequence of a pseudo chaotic random numbers of size N using the Logistic Function described in Chapter 3,say, $\{Y_i: i=1,2,3,\ldots N\}$.
**Step4**: Generate a stream of binary bits, say $\{Z_n: n=1, 2, 3,\ldots, N\}$, by applying the threesholding technique.
**Step5**: Apply the XOR operation between both the generated binary bits in step1 and Step 4, to get a stream of binary bits $\{P_n: n=1, 2, 3,\ldots, N\}$
$P_n = X_n \text{ xor } Z_n$
**Step 6:** Convert the resultant stream of binary bits in step 5 to the correspond characters to form the recovered plaintext (secret message). Fig 7. shows the block diagram of the methodology of the *CryptSteg* system

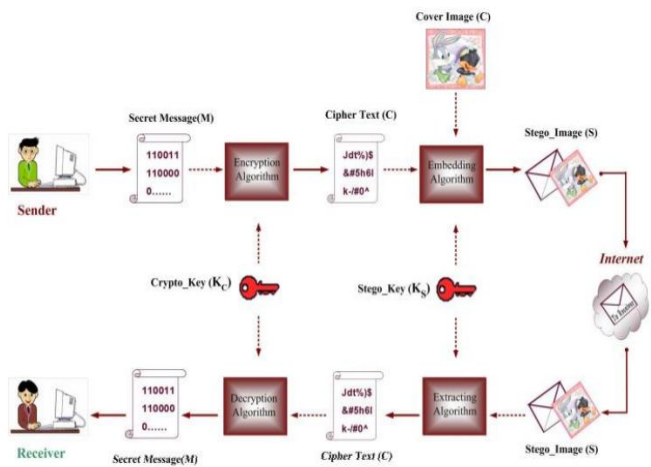

**Figure 7:**. The Methodology of CryptSteg System [23]

### 3.2 Security of the CryptSteg System

The security of the developed system for hiding an encrypted data in a cover image lies in the following main points:
1- The system combines private-key cryptography and the private-key steganography techniques in one system, i.e. the secret message is encrypted using the chaotic cryptography technique and then the encrypted message is embedded in the *cover image*
2- The chaotic encryption/decryption is of type symmetric that is used a secret key (crypto-*key*), $K_c$, shared between the sender and the receiver. This key is one-time password, i.e. it is changed each time the system is running.
3- Similarly, the embedding/extracting processes depending on a secret *key (Stego-key), $K_s$*. shared between the sender and the receiver. This key is one-time password, i.e. it is changed each time the system is running.
4- The embedding process is depending on random choosing of the *cover image* pixels that are candidate for embedding the binary bits of the encrypted secret message.
5- The *stego-key*, $K_s$, depend secret message size, as the secret message change the *stego-key*, as well, will change.
6- There are a data base of cover images so it will change each time the sender(Alice) going to send secret data to the receiver (Bob), this is to remove the suspicion of the Eve when the same image is using frequently.

## 4. CONCLUSIONS AND FUTURE WORK

This paper proposed a system which combines schemes of cryptography with steganography for hiding secret messages and to add more complexity for steganalysis. The proposed system encoded the secret message using chaotic stream cipher and afterwards the encoded data is hidden behind an RGB or Gray cover image by modifying the kth least significant bits (k-LSB) of cover image pixels. The resultant stego-image less distorters. However, to maximum confusion and diffusion of the encrypted image, the computation time of the encryption nad decryption should be enhanced.